\documentclass[12pt, epsfig]{article}
\usepackage{epsfig}
\usepackage{amssymb}
\usepackage{amsfonts}
\newskip\humongous \humongous=0pt plus 1000pt minus 1000pt

\newif\ifdtup

\def\baselinestretch{1.2}
\catcode`\@=11

\@addtoreset{equation}{section}
\def\theequation{\thesection\arabic{equation}}

\def\@normalsize{\@setsize\normalsize{15pt}\xiipt\@xiipt
\abovedisplayskip 14pt plus3pt minus3pt%
\belowdisplayskip \abovedisplayskip
\abovedisplayshortskip \z@ plus3pt%
\belowdisplayshortskip 7pt plus3.5pt minus0pt}

\def\small{\@setsize\small{13.6pt}\xipt\@xipt
\abovedisplayskip 13pt plus3pt minus3pt%
\belowdisplayskip \abovedisplayskip
\abovedisplayshortskip \z@ plus3pt%
\belowdisplayshortskip 7pt plus3.5pt minus0pt
\def\@listi{\parsep 4.5pt plus 2pt minus 1pt
     \itemsep \parsep
     \topsep 9pt plus 3pt minus 3pt}}

\relax

\catcode`@=12

\catcode`\@=11

\def\section{\@startsection{section}{1}{\z@}{3.5ex plus 1ex minus
   .2ex}{2.3ex plus .2ex}{\large\bf}}

\def\thesection{\arabic{section}.}

\def\appendix{\setcounter{section}{0}
 \def\thesection{Appendix \Alph{section}:}
 \def\theequation{\Alph{section}.\arabic{equation}}}

\begin{document}

\newcommand{\beq}{\begin{equation}}
\newcommand{\eeq}{\end{equation}}
\newcommand{\bea}{\begin{eqnarray}}
\newcommand{\eea}{\end{eqnarray}}
\newcommand{\beas}{\begin{eqnarray*}}
\newcommand{\eeas}{\end{eqnarray*}}
\newcommand{\defi}{\stackrel{\rm def}{=}}
\newcommand{\non}{\nonumber}
\newcommand{\bquo}{\begin{quote}}
\newcommand{\enqu}{\end{quote}}
\newcommand{\p}{\partial}
\def\de{\partial}
\def\Tr{ \hbox{\rm Tr}}
\def\const{\hbox {\rm const.}}
\def\o{\over}
\def\im{\hbox{\rm Im}}
\def\re{\hbox{\rm Re}}
\def\bra{\langle}\def\ket{\rangle}
\def\Arg{\hbox {\rm Arg}}
\def\Re{\hbox {\rm Re}}
\def\Im{\hbox {\rm Im}}
\def\diag{\hbox{\rm diag}}
\def\stroke{\vrule height8pt width0.4pt depth-0.1pt}
\def\topfleck{\vrule height8pt width0.5pt depth-5.9pt}
\def\botfleck{\vrule height2pt width0.5pt depth0.1pt}
\def\Zmath{\vcenter{\hbox{\numbers\rlap{\rlap{Z}\kern 0.8pt\topfleck}\kern
2.2p \rlap Z\kern 6pt\botfleck\kern 1pt}}}
\def\Qmath{\vcenter{\hbox{\upright\rlap{\rlap{Q}\kern
                   3.8pt\stroke}\phantom{Q}}}}
\def\Nmath{\vcenter{\hbox{\upright\rlap{I}\kern 1.7pt N}}}
\def\Cmath{\vcenter{\hbox{\upright\rlap{\rlap{C}\kern
                   3.8pt\stroke}\phantom{C}}}}
\def\Rmath{\vcenter{\hbox{\upright\rlap{I}\kern 1.7pt R}}}
\def\Z{\ifmmode\Zmath\else$\Zmath$\fi}
\def\Q{\ifmmode\Qmath\else$\Qmath$\fi}
\def\N{\ifmmode\Nmath\else$\Nmath$\fi}
\def\C{\ifmmode\Cmath\else$\Cmath$\fi}
\def\R{\ifmmode\Rmath\else$\Rmath$\fi}
\def\QATOPD#1#2#3#4{{#3 \atopwithdelims#1#2 #4}}
\def\stackunder#1#2{\mathrel{\mathop{#2}\limits_{#1}}}
\def\stackreb#1#2{\mathrel{\mathop{#2}\limits_{#1}}}
\def\Tr{{\rm Tr}}
\def\res{{\rm res}}
\def\Bf#1{\mbox{\boldmath $#1$}}
\def\balpha{{\Bf\alpha}}
\def\bbeta{{\Bf\beta}}
\def\bgamma{{\Bf\gamma}}
\def\bnu{{\Bf\nu}}
\def\bmu{{\Bf\mu}}
\def\bphi{{\Bf\phi}}
\def\bPhi{{\Bf\Phi}}
\def\bomega{{\Bf\omega}}
\def\blambda{{\Bf\lambda}}
\def\brho{{\Bf\rho}}
\def\bsigma{{\bfit\sigma}}
\def\bxi{{\Bf\xi}}
\def\bbeta{{\Bf\eta}}
\def\d{\partial}
\def\der#1#2{\frac{\d{#1}}{\d{#2}}}
\def\Im{{\rm Im}}
\def\Re{{\rm Re}}
\def\rank{{\rm rank}}
\def\diag{{\rm diag}}
\def\2{{1\over 2}}
\def\ntwo{${\cal N}=2\;$}
\def\4N{${\cal N}=4$}
\def\none{${\cal N}=1\;$}
\def\x{\stackrel{\otimes}{,}}
\def\beq{\begin{equation}}
\def\eeq{\end{equation}}
\def\ba{\beq\new\begin{array}{c}}
\def\ea{\end{array}\eeq}
\def\be{\ba}
\def\ee{\ea}
\def\stackreb#1#2{\mathrel{\mathop{#2}\limits_{#1}}}
\def\baselinestretch{1.0}

\begin{titlepage}

{\hfill IFUP-TH/2003-23,   ITEP-TH-42/03, TIT-HEP/506 }  
\bigskip
\bigskip
\bigskip
\bigskip

\begin{center}
{\Large {\bf
  NONABELIAN SUPERCONDUCTORS:  VORTICES  AND CONFINEMENT \\
IN \ntwo SQCD
 } }
\end{center}
\vspace{1em}

\begin{center}
{\large  Roberto AUZZI $^{(1,3)}$ , Stefano BOLOGNESI $^{(1,3)}$, \\
 Jarah EVSLIN $^{(3,2)}$, Kenichi KONISHI $^{(2,3,4)}$,    Alexei YUNG $^{(5,3,6)}$
 \vskip 0.10cm
 }
\end{center}

\begin{center}
{\it   \footnotesize
Scuola Normale Superiore - Pisa $^{(1)}$,
 Piazza dei Cavalieri 7, Pisa, Italy \\
Dipartimento di Fisica ``E. Fermi" -- Universit\`a di Pisa $^{(2)}$, \\
Istituto Nazionale di Fisica Nucleare -- Sezione di Pisa $^{(3)}$, \\
     Via Buonarroti, 2, Ed. C, 56127 Pisa,  Italy $^{(2,3)}$ \\
Dept. of Physics, Tokyo Inst.  of Technology $^{(4)}$,
                     2-12-1 Oh-okayama, Meguro-ku,
            Tokyo 152-8551 Japan \\
Petersburg Nuclear Physics Institute $^{(5)}$, 
            Gatchina,
            188300 St. Petersburg, Russia \\
Institute of Theoretical and Experimental Physics $^{(6)}$, 
B. Cheryomushkinskaia 25, 117259 Moscow, Russia\\
   }

\end {center}

\vspace{0.5  em}

\noindent
{\bf Abstract:}

  We  study nonabelian vortices (flux tubes) in $SU(N)$ gauge theories, which are responsible for the confinement of  (nonabelian)   magnetic monopoles.
In particular   a detailed analysis is given    of
 \ntwo SQCD with gauge group $SU(3)$     deformed by a small  adjoint
chiral multiplet mass.      Tuning the   bare quark masses (which we take to be large) to a common value $m$, we consider
a particular  vacuum of this theory in which an $SU(2)$   subgroup of the
gauge group remains unbroken.    We consider  $5
\ge  N_f
\ge 4$ flavors   so that  the   $SU(2)$ sub-sector  remains  non asymptotically free: the vortices carrying nonabelian fluxes may be  reliably  studied
in a semi-classical regime.
 We show that  the
vortices indeed acquire  exact    zero modes which generate   global rotations of the flux in  an  $SU(2)_{C+F}$   group.
   We study an effective
world sheet theory of these  orientational zero modes which  reduces to an \ntwo $O(3)$ sigma model in (1+1) dimensions.
Mirror symmetry then teaches us  that  the dual $SU(2)$  group is not dynamically broken.

\vfill

\begin{flushright}
July    2003

\end{flushright}

\end{titlepage}

\bigskip

\hfill{}
\bigskip

\section{Introduction and Discussion }

         Some sort of nonabelian vortices  are believed to be responsible for
 confinement in QCD.
Although in string theory these objects appear naturally, they turn
out to be somewhat elusive in four-dimensional
field theories. The existing literature on the subject certainly provides an
 incomplete picture.

There are several reasons for this unsatisfactory situation.
One of the reasons is that  boundstates of vortices are
not generally stable.
An example is the
case of an $SU(N)/{\mathbb Z}_N$ gauge theory (e.g., $SU(N) $
  gauge theory with all fields in the adjoint representation)
 broken completely by a Higgs mechanism, where possible vortices
represent nontrivial elements of the fundamental group \beq
\label{center}
\pi_1( SU(N)/{\mathbb Z}_N )= {\mathbb Z}_N.
\eeq
${\mathbb Z}_N$-charged objects cannot be BPS saturated \cite{HSZ, KS},   and this fact, together
 with the unknown dependence of their properties on the form
of the potential, number of the fields, etc., has obstructed investigations of such vortices.

Secondly, often these theories  become  strongly coupled at low energies and
therefore an analytical study of the vortex configurations is very
 difficult.
For instance, confinement in QCD  may  be due to the  vortices of  electric fields
appearing in a dual (magnetic) ($SU(3),  SU(2) \times U(1),$  or  $U(1)^2$?)
theory. Unfortunately, neither the true nature of the effective magnetic
degrees of freedom nor their form of interactions is
known at the moment. 't Hooft's suggestion that they be abelian monopoles   of
a gauge-fixed $U(1)\times U(1)$ theory \cite{TM},
must still be verified. On the other hand, there is no experimental
indication that  the $SU(3)$ gauge group is dynamically broken
to $U(1)\times U(1)$.

Finally, in  the examples of classical solutions for ``nonabelian vortices"
discussed   so far
in the literature \cite{Vtx}          the vortex flux is actually
always oriented in a fixed direction in the Cartan subalgebra,  showing  that   they   are basically  abelian.

Useful  hints  come from
the  detailed   study of  a  wide class of   softly broken \ntwo super\-symmetric gauge theories where the dynamics appears  particularly
transparent. It was  shown that,  in fact,  different types of confining vacua are  realized in these models \cite{SW1,SW2,curves,DS,CKM}.            It
is possible that in some cases  confinement is due to the condensation of monopoles associated with the maximally abelian subgroup (a dual Meissner
effect), as in the
\none vacua surviving the adjoint mass perturbation in the pure
\ntwo SYM
\cite{SW1,curves,DS}. These cases  provided the first examples of four-dimensional
 gauge
theory models in which the 't Hooft-Mandelstam mechanism of confinement \cite{TM}  is    realized and can be analysed quantitatively.
A detailed study of these cases has  shown however   that {\it  dynamical abelianization}     takes place there, with
a  characteristically richer meson
spectrum
\cite{DS,HSZ,S,VY}.
Indeed the low-energy effective gauge group of the   $SU(N)$ theory is  $U(1)^{N-1}$ and
the meson spectrum is
classified according to the number of possible abelian strings via
\beq
\label{zn}
\pi_1( U(1)^{N-1} )= {\mathbb Z}^{N-1},
\eeq
({\it cfr.} (\ref{center})).   Thus vorties and therefore mesons come in infinite towers,   a feature
not expected in the real world  QCD.

However, such  is not    the typical situation in softly broken
\ntwo theories with fundamental matter fields (quarks) \cite{APS,CKM}.
{\it    Confining vacua in $SU(N)$, $SO(N)$ and $USp(2N)$ gauge
theories with $N_f$ quark flavor, are typically described by
effective  nonabelian    dual gauge theories.}   For instance,
 in the so-called $r$-vacua of $SU(N)$ gauge theory with
$N_f$ flavors and vanishing bare quark masses, the
low-energy effective theory is a dual $ SU(r)\times U(1)^{N-r} $ theory.
Addition of the adjoint chiral multiplet mass term
$\mu \, \Tr \, \Phi^2$ breaks supersymmetry to ${\cal N}=1$, and the dual quarks in
the ${\underline r}$ of $SU(r)$ condense.
These ``dual
quarks" have been recently identified  \cite{BK}  as  the
quantum Goddard-Nuyts-Olive-Weinberg monopoles \cite{GNO,EW}.
Their condensation is believed to
give rise to  nonabelian confinement
via formation of nonabelian flux tubes.

In fact, the problem of nonabelian vortices is very closely related to
(in a sense, it is one and the same problem as)
that of the nonabelian monopoles on which they end.
A key feature found in   \cite{BK}   is that the quantum behavior of the nonabelian monopoles,  and in fact the vacuum
properties themselves   depend critically on the presence of massless flavors of matter.  We shall find below    that the existence of
   nonabelian  {\it  vortices} similarly   requires     the presence of    massless  flavors in the  underlying theory.

Inspired by these developments, and based on a   work by Marshakov and one of the authors (A.Y.)
\cite{MY},     we   present   in this paper  a  study of    nonabelian superconductors,  concentrating  our attention on the
properties of the vortices appearing in these systems.  In a companion paper \cite{five}, we shall explore      more extensively the properties of
nonabelian monopoles
themselves.

 Our analyses are done
 in a context where the dynamics of the model
is well understood and the transition from a theory with  abelian vortices  to one with  nonabelian
vortices can be studied in a weakly coupled semi-classical
regime throughout.
The model we consider   is probably the simplest of such models, \ntwo QCD with gauge group
$SU(N)$ and $N_f$ hypermultiplets of fundamental matter (quarks).
Upon deformation of this theory via a small  mass term   for the adjoint
chiral multiplet,  $ \mu \, \Tr \Phi^2$,      the Coulomb branch of the theory shrinks to a number of isolated
\none vacua.

Generically the vacuum expectation value (VEV) of the adjoint field breaks the
$SU(N)$ gauge symmetry down to $U(1)^{N-1}$.
However,
it was shown in \cite{APS,CKM} (see also \cite{MY})
that some of the \none
vacua of $SU(N)$ \ntwo QCD   preserve a nonabelian subgroup.
These vacua are classified by an integer  $ r$.
 In  a semiclassical regime,  which is valid  at
 large bare  quark masses
\beq   m_A \gg \Lambda, \qquad      A=1,\ldots N_f, \eeq         the adjoint scalar VEVs   in those vacua   take 
the form,
\beq   \bra \phi \ket = { 1\o \sqrt 2}     \,  \diag  (- m_1,  - m_2, \ldots, -  m_r,   c,  c, \ldots,  c),  \qquad  c=  { 1\o N-r} \sum_{k=1}^r m_k,
\label{Advev}\eeq
where   $r$ quark masses out of $N_f$  possible masses are chosen to satisfy   the
vacuum equations.

When  the  quark masses are tuned to a common  value $m$,  the pattern of the spontaneous breaking changes to
\beq   SU(N) \to  SU(r)  \times SU(N-r) \times U(1).
\eeq
   The $SU(N-r)$ sector   is a pure \ntwo  Yang Mills theory
\footnote{Recall that the quark masses come from the superpotentials
$  {\tilde Q}_i   ( \sqrt 2 \phi + m_i )  Q_i.    $}  and   becomes strongly interacting at low energies and gets
dynamically broken  to $U(1)^{N-r-1}$.    The $SU(r)$  sector, on the other hand,
having
$N_f$ massless  flavors, remains weakly coupled    as long as   $r \le    { N_f \o 2}$.

  Furthermore,  in the presence of the aforementioned adjoint mass perturbation, the  light  squark fields acquire
VEVs of     color-flavor diagonal form  (``Color-Flavor Locking"),
\beq      \bra  q_i^a  \ket   =   \delta_{i}^{a}   \,\sqrt{\mu \,  m}, \qquad  i,a =1,2, \ldots r;
\label{Qvev}\eeq
which breaks the $SU(r) \times U(1) $ gauge group completely at scales far below the bare quark masses:  $\sqrt{\mu m} \ll m$.
The theory is now  in the Higgs phase, and develops vortex configurations, representing nontrivial elements of
\beq \pi_1 ( {SU(r) \times U(1)^{N-r}  \o {\mathbb Z}_r} )  =  {\mathbb Z}^{N-r} .   \eeq
The key fact is  that  the system has an exact global  $SU(r)_{C+F}$ symmetry, respected both by the interactions and by the scalar VEVS
(\ref{Advev}) and (\ref{Qvev}).  A given vortex configuration however  breaks  this symmetry:   it turns out that the   symmetry is broken as
$SU(r) \to SU(r-1) \times U(1)$ (see  below.)  As a result,
exact orientation  zero modes of   $SU(r) /  ( SU(r-1) \times U(1) )  \sim  {\bf CP}^{r-1}$  are generated.

To work things out  concretely,       we analyse    the case of the $ r=2$ vacua of the
$SU(3)$  gauge theory  ($N=3, $  $r=2$  above)   in detail in the main body of this  paper.
   The
$SU(2)$ subgroup, classically restored in the limit of equal quark masses,
stays unbroken   in the full quantum theory,   as the relevant sector of the theory is infrared free   if    $N_f > 4$,
  or    is conformal invariant  if  $N_f=4$ .
On the other hand,  of course, the underlying  $SU(3)$ gauge theory
is   asymptotically free  for  $N_f \le  5$, so we shall  take $N_f$ to be either $4$ or $5$.

This is one of the  important   points     of our analysis:
by working in the regime in which   the  interactions   remain  weak at all scales,      the continuous    transition from the  theory with  abelian vortices  (unequal quark
masses)
to the theory  with
 nonabelian vortices   which are qualitatively different,    can be studied  explicitly and    reliably.

The unbroken gauge group $SU(2) \times U(1) / {\mathbb Z}_2$  is further broken at a much lower mass scale, yielding vortices
representing the  nontrivial elements of 
\beq \pi_1 ( {SU(2) \times U(1) \o {\mathbb Z}_2} )  =  {\mathbb Z}.   \eeq
Indeed, as  the bare  quark masses are tuned to a common value,    $ m_i \to m$,    starting from unequal and generic values,
 the low-energy gauge group gets enhanced
from $U(1)\times U(1)$ to $SU(2)\times U(1)$.    The set of    abelian vortices  appearing in the unequal mass cases
acquires a  certain degeneracy  and   at the same time some   orientation  (in the color space) zero modes  appear    which relate
the vortices  of the same tension by   global       rotations.     These zero
modes are associated with the diagonal global $SU(2)_{C+F}$ subgroup
   of color  $SU(2)_C$ crossed with the flavor  $   SU(2)_F \subset SU(4)_F$ which   is an exact symmetry of the system.   More precisely, the vortex zero modes
parametrize $SU(2) / U(1)  \sim {\bf CP}^1  \sim S^2$  as  each  vortex solution breaks the exact $SU(2)_{C+F}$  symmetry to a $U(1)$ subgroup.

We then  work out the effective  world-sheet theory of  the  vortex
zero modes,    and show that it reduces to the \ntwo $O(3)$ sigma model in
(1+1) dimensions. Classically    the   $O(3)$ sigma model has spontaneous symmetry
breaking and appears to  yield massless Goldstone fields. In terms of strings in four
dimensions this would  mean that $SU(2)_{C+F}$ is spontaneously broken
and the string flux is oriented in some particular direction inside
the $SU(2)_C$ gauge subgroup.

However  the  quantum physics of the \ntwo $O(3)$ sigma model in (1+1) is
quite different. It is well understood using the  mirror
map  \cite{HoVa}, which relates it to a sine-Gordon theory.
In particular it is known   that the model has a mass gap and no spontaneous symmetry
breaking. In terms of strings in 4D this means that the string is
not oriented in any particular direction inside $SU(2)_C$ group.
This ensures   that   our vortices are  truly nonabelian.
The sine-Gordon superpotential is generated dynamically in the
effective (1+1)-dimensional worldsheet theory  
  which produces  exactly
two   vacua.

Our considerations can be straightforwardly generalized to the    $r= N-1 $  vacua  of the $SU(N)$  theory  with $2 N \ge  N_f \ge   2 (N-1)$, 
 with unbroken  $SU(N-1)$  group, although our analysis in these more general cases is less complete.   In particular, in the case of an
$SU(N)$ theory broken to
$SU(N-1)\times U(1)$ the zero modes of the vortex are described by a 2-dimensional {\bf CP}${}^{N-2}$ sigma model whose mirror is an affine Toda theory
with the desired
$N-1$ vacua.

 The vortices studied in this paper, though    stable in the low-energy theory,  are   strictly speaking
 metastable       as   the underlying  gauge group  ({\it e.g.,} $SU(3)$)    is simply connected.    Their
decay rates are however     small,     being exponentially suppressed by ratios of heavy
monopole  masses   squared to the string tensions \cite{Vi,SY}  .

Our result  provides, albeit indirectly, a couterexample  to the no-go theorem on the existence of  monopoles with nonabelian
charges discussed earlier \cite{CDyons}.   These  nonabelian monopoles do exist in our theory as stable solitons and act   as the sources of the
nonabelian vortices considered  here,    and are actually   confined by    them.    We exhibit here  explicitly the
transformations among the vortices,  which imply   certain non-local transformations for their sources. We will see that the zero modes of the vortices are
normalizable.  To calculate the zero mode of a single monopole, which necessarily sources an infinite vortex, we must integrate that of the vortex along its infinite length.
Thus we find, as was seen in the flavorless cases of Refs.~\cite{CDyons}, that the zero mode of a single monopole is nonnormalizable.  In a color-neutral configuration of
monopoles the total length of the vortices may be taken to be finite and so the integral is finite, yielding normalizable zero modes which again generalize those known to
exist in the flavorless case.

Throughtout this paper we limit ourselves to cases with   large bare quark masses
where the original electric subgroup remains weakly coupled.   When the bare quark masses are tuned to small values or even to
zero,  the low-energy  system  is weakly coupled when described
in terms of 
the magnetic variables instead of  the electric ones.   The
excitations which are quarks in the electric description  
at large quark masses become monopoles in the magnetic
description at small quark masses \cite{SW2,CV}. 
The properties of the corresponding $r$- vacua have been studied
in detail in \cite{CKM},  and in the case of  a SCFT $r=2$ vacua
 of $SU(3)$ theory,  in \cite{AGK}.     The properties of these
corresponding vacua are closely related by holomorphy.

The organization of the paper is as follows. In Sect. 2 we
review \ntwo QCD with  equal quark masses,
 work out its low-energy description, vacuum structure and the
low-energy spectrum. In Sect. 3 we derive nonabelian Bogomolny equations, construct  vortices and   study  their
$SU(2)$   zero modes.   We discuss the generalization to the more general case of $SU(N) \to  SU(N-1)  \times U(1)$
breaking  in Sect. 4.   In Sect. 5 we work out the effective world sheet theory for orientational zero modes and
discuss its physics.

While this work was in preparation   Ref.~\cite{HT} appeared which considers vortices in the very similar \ntwo three-dimensional theory with an FI term. 
While these vortices are not strings but particles, the worldvolume theories appear to be related by dimensional reduction, and the vacuum structures and
spectra appear to be the same.  Thus many of our results as well as an extensive analysis of the relevent moduli spaces may be found there.

\section{\ntwo $SU(3)$ QCD}
\setcounter{equation}{0}
\label{model}

\subsection{The Model }

The field content
of \ntwo QCD with the gauge group SU(3) and $N_f$ flavors of chiral
multiplets is as follows. The \ntwo vector multiplet consists of the
gauge field $A_{\mu}$, two Weyl fermions $\lambda^{1}_{\alpha}$,
 $\lambda^{2}_{\alpha}$ and
the scalar field $\phi$, all in the adjoint representation of the
gauge group. Here   $\alpha=1,2$ is
a spinor index   while all adjoint fields
 are $3\times 3$ matrices in the Lie algebra $SU(3)$.

The chiral multiplets of the $SU(3)$ theory consist of   complex
 scalar squarks
$q^{k A}$ and $\tilde{q}_{A k }$ and   Weyl fermion quarks $\psi^{k   A}$ and
$\tilde{\psi}_{A k}$,
 all in the fundamental representation of the gauge group.
Here
 $k =1,2,3$ is a color index
while $A$ is a flavor index, $A=1,\ldots N_f$.

This theory has a Coulomb
branch on which the adjoint scalar acquires the vacuum expectation value (VEV)
\beq
\label{phigen}
\phi =
\frac{1}{2}\left(
\begin{array}{ccc}
  a_3+\frac{a_8}{\sqrt{3}} & 0 & 0 \\
  0 & -a_3+\frac{a_8}{\sqrt{3}} & 0 \\
  0 & 0 & -2\frac{a_8}{\sqrt{3}}
\end{array}\right) \equiv \lambda_3a_3 + \lambda_8a_8,
\eeq
generically breaking the
 $SU(3)$  gauge group down to  $U(1)\times U(1)$. Here
 $\lambda_3$ and $\lambda_8$ are the Gell-Mann matrices
of the Cartan subalgebra.

In this paper  we consider the special vacua for which
\beq
\label{a3vev}
<a_3>=0.
\eeq
For these vacua the low-energy gauge group is $SU(2)\times U(1)$, at least
classically.

 We perturb the above theory by adding a small
mass term for the adjoint matter  via the superpotential
\beq
\label{brsup}
{\cal W}=\mu\; \Tr \, \Phi^2\,.
\eeq

Generally speaking, the superpotential breaks
\ntwo down to ${\cal N}=1$.
The Coulomb branch shrinks to
a number of  isolated \none vacua \cite{APS,CKM}.
In the limit $\mu\to 0$ these vacua correspond to special
singular points on the Coulomb branch
in which  pairs of monopoles/dyons or
quarks become massless.
Three of these points are always at
strong coupling. They correspond to \none vacua of the pure
gauge theory.
The massless quark points are   at weak
coupling if   the quark masses $m_A$ are large,
$m_A\gg \Lambda$.  The vacua
in which quarks become massless will be referred to as the quark vacua.
We shall  be mainly interested in these quark vacua.

It is   important to note  that
 \ntwo supersymmetry is  not
broken  to the leading order in the parameter $\mu$
in the effective theory \cite{HSZ,VY}.
In the effective low-energy  theory the superpotential
(\ref{brsup}) gives rise to a superpotential linear in  $a_8$ plus
 higher order corrections.
If only the linear term  in $a$'s in the superpotential
is kept   and if we restrict our attention to the special vacua (\ref{a3vev}),
then it reduces to a \ntwo Fayet-Iliopoulos term which
does not break the  \ntwo supersymmetry.

\subsection{ $SU(2)\times U(1)$ symmetric  low-energy  theory }

The $SU(3)$ gauge
group is broken down to $U(1)\times U(1)$ by the VEV of the adjoint scalar
(\ref{phigen}) at generic values of  quark masses. However,   in the equal quark
mass limit ($m_A=m$)
which we shall consider from now on,      the VEV of the $a_3$ field vanishes   (see Sec.~(\ref{sec:vacua})), and       the low-energy gauge group is $SU(2)\times
U(1)$. W-bosons which are charged with respect to both factors of the low-energy group acquire a large mass of order $m$. The third color
components of quarks also become heavy
in   this  vacua  with masses of order of $m$.

Let us  consider now  the scales of order $\sqrt{\mu \,  m}$, which are
well below W-boson masses ($\mu$ is taken small,    $\mu\ll m$).  There
the low-energy theory contains the following light fields
of the \ntwo vector multiplet:
 four complex  scalar light fields $a_b$ and $a_8$
where $b =1,2,3$ is the color $SU(2)$ index,  one $SU(2)$
gauge field $A^{b}_{\mu}$ and one
$U(1)$ gauge field
 $A^{(8)}_{\mu}$ together with their fermionic superpartners.
For example the gauge fields are defined as follows:
\beq
\label{Amu}
A_{\mu} =
 \lambda_b \, A^{b}_{\mu}+
\lambda_8 \,  A^{8}_{\mu}
\eeq
where our notation corresponds to expanding  gauge and adjoint fields
in the orthogonal  basis of the Gell-Mann matrices, $\lambda_a$
being the  first
three Gell-Mann matrices normalized as
 $Tr(\lambda_a \lambda_b)=1/2\;\delta_{ab}.$

Light quark multiplets contain complex scalar $SU(2)$-doublets
$q^{kA}$, $\tilde{q}_{Ak}$ together with their fermionic superpartners, $k=1,2$.

The bosonic part of the low-energy effective
 theory then  acquires the form
\bea
S_{eff}=   \int d^4x \!\! \,&\Big[ &   \frac1{4g^2_2}\left(F^{b}_{\mu\nu}\right)^2 +
\frac1{4g^2_1}\left(F^{8}_{\mu\nu}\right)^2
+  \frac1{g^2_2}\left|D_{\mu}a_b\right|^2 +\frac1{g^2_1}
\left|\partial_{\mu}a_8\right|^2
\non \\ 
&+ &   \left|\nabla_{\mu}
q^{A}\right|^2 + \left|\nabla_{\mu} \bar{\tilde{q}}^{A}\right|^2
+V(q^A,\tilde{q}_A,a_b,a_8)     \, \Big] .
\label{qed}
\eea
Here $D_{\mu}$ is the covariant derivative in the adjoint representation
of $SU(2)$ gauge subgroup,
while
\beq
\label{nabla}
\nabla_\mu=\partial_\mu -\frac{i}{2\sqrt{3}}\; A^{8}_{\mu}
-i A^{b}_{\mu}\frac{\tau^b}{2},
\eeq
where we suppress the color $SU(2)$ indices and $\tau^b$ are
$SU(2)$ Pauli matrices. The coupling constants $g_1$ and $g_2$ correspond
to $U(1)$ and $SU(2)$ sectors respectively.
The potential in the Lagrangian (\ref{qed})
is given by the   D and F terms
$$
V(q^A,\tilde{q}_A,a_b,a_8)=
 \frac{g^2_2}{8}
\left(\bar{q}_A  \,  \tau^b q^A - \tilde{q}_A\tau^b \bar{\tilde{q}}^A\right)^2
 + \frac{g^2_1}{24}
\left(\bar{q}_A \,q^A - \tilde{q}_A \, \bar{\tilde{q}}^A\right)^2+
$$
\beq 
+ \frac{g^2_2}{2}\left|\, \tilde{q}_A \,  \tau^b q^A\right|^2+
\frac{g^2_1}{6}\left| \, \tilde{q}_A \, q^A+
 \sqrt{6}\, \mu <a_8>\right|^2 + \ldots, 
\label{pot}\eeq  
where other D-terms involving the  adjoint scalar fields $a_8$ and $a_b$    ($b=1,2,3$) 
(which vanish at  $\bra a_8 \ket \ne 0 $ and   $\bra a_b \ket  = 0 $) are  left implicit.   
The term $\sqrt{6}\, \mu <a_8>$ in the second line arises when we expand
fields $a_8$ and $a_b$     in the
superpotential (\ref{brsup})
around their VEV's  and keep only terms  linear in
fluctuations of these fields.
As we have   already noted,   this means that the theory in (\ref{qed})
is a bosonic part of a \ntwo supersymmetric theory. In particular
this ensures that our theory has BPS vortices \cite{HSZ,VY,MY}
(see also the seventh ref. in \cite{Vtx}).

The theory (\ref{qed}), (\ref{pot}) is an  $SU(2)\times U(1)$ generalization
of the low-energy theory for the $U(1)\times U(1)$ case studied  in
\cite{MY}.

Below the scale $m$ the $SU(3)$ gauge group is broken and we have
two coupling constants $g_1$ and $g_2$   which run
according to the $U(1)$ and $SU(2)$ renormalization
group flows respectively. Note
 that with a logarithmic accuracy we can neglect mixing of these two
coupling constants. In the case with   four flavors     the $SU(2)$ coupling
does not run ($SU(2)$ theory with $N_f=4$ is conformal) and is given
by its value at the scale $m$
\beq
\label{g2}
\frac{8\pi^2}{ g^2_2}= 2\log{\frac{m}{\Lambda}}+\cdots.
\eeq
Since at large $m$ the $SU(2)$ sector is weakly coupled, it remains so at low energies.

The $U(1)$ coupling
undergoes an additional renormalization from the scale $m$ to the
scale determined by the masses of light states in the low-energy theory
(which  are  of the  order of $\sqrt{\mu m}$, see next subsection). Thus we have
\beq
\label{g1}
\frac{8\pi^2}{ g^2_1}= 2\log{\frac{m}{\Lambda}}+
 \frac23\log{\sqrt{\frac{m}{\mu}}} +\cdots,
\eeq
where we use the fact that the one loop coefficient of the $\beta$-function
for $U(1)$ theory is $~b=- 2\,n_e \,  N_f$ and substitute $N_f=4$ and the
electric charge $n_e=1/2\sqrt{3}$, see (\ref{nabla}). Clearly,
this coupling is even smaller than the one in the $SU(2)$ sector.

If the number of the quark flavors is taken to be five, the $SU(2)$ gauge coupling constant also runs
to smaller values  towards  the infrared.   In general therefore one has $g_1 \ne g_2$, both small, and we shall not need
more details in the analyses below.

\subsection{Vacuum structure and low-energy spectrum   \label{sec:vacua} }

In this subsection we review the vacuum structure and low-energy mass spectrum of $SU(3)$ \ntwo QCD \cite{APS,CKM}
generalizing the analysis made in \cite{MY} to the case of the
$SU(2)\times U(1)$ low-energy group.
To find the   vacua of the effective theory (\ref{qed}) we have to look for
the zeros of the potential (\ref{pot}). At generic
large values of quark masses  solutions
have the following structure \cite{CKM,MY}. Besides the three strong
coupling vacua which exist already in the pure $SU(3)$ gauge theory
there are $2N_f$  $r=1$ vacua and $N_f(N_f-1)/2$ $r=2$  vacua,
were $r$ is the number of quark flavors which develop non-zero VEV's.

Here we are mostly interested in $r=2$ vacua, which have an  $SU(2)\subset SU(3) $  unbroken gauge group which becomes exact
in the case of equal quark masses.
Clearly the minimal number of flavors for which we can have a
$r=2$ vacuum is $N_f=2$. Let us  consider this case first.

The adjoint scalar matrix is given by
\beq
\label{phi}
\phi = -{1\over\sqrt{2}}\left(
\begin{array}{ccc}
  m & 0 & 0 \\
  0 & m & 0 \\
  0 & 0 & -2m
\end{array}\right)
\eeq
 where $m$  is the common mass of both flavors.
 In the above notation (\ref{phi}) reads
\beq
\label{avev}
\langle a_3\rangle = 0,\qquad 
\langle a_8\rangle =-\sqrt{6} \,  m.
\eeq

For real values of $m$  and  $\mu$ we can use gauge rotations to
make squark VEV's real. We write the squark field as a $2\times 2$
matrix $q^{kA}$ where  $k=1,2$ is a color index and
$A=1,2$ is a flavor one. Then the  squark VEV's  are given by
\beq
\label{qvev}
<q^{kA}>=<\bar{\tilde{q}}^{kA}>=\sqrt{\frac{\xi}{2}}\left(
\begin{array}{cc}
  1 & 0  \\
  0 & 1  \\
  \end{array}\right),   \eeq
where we have used  color-flavor mixed  matrix notation, and   we have introduced
\beq
\label{xi}
\xi=6 \,  \mu \, m,
\eeq
which acts as  the  Fayet-Iliopoulos parameter  of the $U(1)$.    $\xi$    sets the scale of the low-energy theory (\ref{qed}).
Only  the  two upper color components and the first two flavors  are  shown in Eq.(\ref{qvev}): all other components
have vanishing VEVS.

Now consider the spectrum of light fields in this vacuum.
The $SU(2)\times U(1)$ low-energy gauge group is broken completely by squark VEV's
and all gauge bosons acquire masses.
The mass matrix for the gauge fields $A^{a}_{\mu}$, $A^{8}_{\mu}$ can be
read off of the kinetic terms for the quarks in (\ref{qed}).
It turns out that it is diagonal in the basis $A^{a}_{\mu}$, $A^{8}_{\mu}$. In particular, the mass
of $A^{8}_{\mu}$ is given by
\beq
\label{m8}
m_{8}^2 = \frac{1}{3}\,g^2_1 \, \xi,
\eeq
while the mass of the $SU(2)$ W-boson is
\beq
\label{m3}
m_{W}^2 = g^2_2 \,  \xi .
\eeq
The masses  of the adjoint scalars   $a_8$ and  $a_b$ are identical
to the ones in (\ref{m8}) and (\ref{m3})  as
can be seen  from (\ref{pot}).

The mass matrix for squarks is now of size $16\times 16$ including four
real components of complex fields $q$ and  $\tilde{q}$ for
each color and flavor.
It has four zero eigenvalues
associated with the four states ``eaten'' by the Higgs mechanism for
$U(1)$  and $SU(2)$ gauge factors and two non-zero eigenvalues coinciding
with gauge boson
masses (\ref{m8}) and (\ref{m3}). The eigenvalue (\ref{m8}) corresponds to
three squark eigenvectors while the one in  (\ref{m3}) corresponds to
nine squark eigenvalues.

Altogether we have one long \ntwo multiplet with mass  (\ref{m8}),
 containing eight bosonic states
(3 states of the massive $A^8_{\mu}$ field plus 2 states of $a_8$
plus 3 squark states)
 and eight fermionic states.
In addition we have three long \ntwo multiplets with mass  (\ref{m3})
labeled by the color index $a=1,2,3$
also containing eight bosonic and eight fermionic states each \footnote{ See \cite{VY} for a discussion of the emergence of \ntwo long
multiplets in Seiberg-Witten theory upon adjoint mass term
deformation.}.
Note that  no Nambu-Goldstone multiplets appear in this vacuum:
 all phases associated with broken symmetries are "eaten" by
Higgs mechanism.

Actually,  in the theory  with $N_f=2$  discussed above, the   $SU(2)$ gauge interactions become
 strong  below the scale $m$, and the properties
of the theory at low energies (at mass scales of order of   $\sqrt{\mu \, m}  \ll m$) cannot be determined
from the Lagrangian (\ref{qed})  only.

For this reason,  we   introduce more flavors into our theory
and consider the  $SU(3)$    theory  with    $N_f=4$ or $N_f=5$.  The low-energy $SU(2)\times U(1)$ then remains in the weak coupling regime.

 This theory has  $ { N_f \choose  2}$   $\,r=2$ vacua
of the type described above,   for unequal quark masses. Each of these vacua corresponds to choosing two flavors out of $N_f$ which develop VEV's. This gives
$N_f(N_f-1)/2 =6$ choices for $N_f=4$. In the limit of equal masses
all six vacua coalesce and a  Higgs branch develops from the common
root. The dimension of this Higgs branch is $8(N_f-2)$ \cite{APS,MY}.
To see this note that we have  $8N_f$ real variables $q^{kA}$
subject to four $D$-term and eight $F$-term conditions in the potential
(\ref{pot}). Also 3+1 gauge phases are eaten by the Higgs mechanism.
Thus we have $8N_f-12-4=8(N_f-2)$ remaining degrees of freedom.

We consider below a special submanifold of the Higgs branch
which admits BPS flux tubes (cf. \cite{HSZ,GS,MY,EY}).
This base submanifold is compact and has the minimal
value of the quark condensate $<|q^A|^2>=<|\tilde{q}_A |^2>=\xi$.
 One point on 
this submanifold  which corresponds to non-zero VEV of the
first flavor and non-zero VEV of the second flavor
while all other components are zero is given in (\ref{qvev}).

Other points on the base of the Higgs branch are given by a $SU(N_f)$
flavor rotation of (\ref{qvev}). The dimension of the base
submanifold  of the
Higgs branch is $4(N_f-2)$ \cite{MY}. To see this note that VEV's of
two flavors  break $SU(N_f)$ symmetry down to
$SU(N_f-2)$. Thus the number of "broken" generators is
$\dim SU(N_f) - \dim SU(N_f-2) = 4(N_f-1)$
and also we have  to
subtract four phases  ``eaten'' by the Higgs mechanism.

Other points on the $8(N_f-2)$ dimensional Higgs
branch correspond to non-zero VEV's of massless moduli fields, and
these points do not admit BPS strings. In particular, the ANO
strings \cite{ANO}   on the Higgs branch were studied in \cite{Y99,EY}, they
correspond to a limiting case of type I strings with the
logarithmically thick tails associated with massless scalar
fields.  We shall    not discuss here strings
at generic points on the  Higgs branch.

Before ending this subsection, we  need to comment on the soliton sector.
  In the  monopole sector,  all  solitonic states associated with the symmetry breaking (\ref{phi})  are massive.   In particular, one finds
an exactly degenerate doublet of BPS  monopoles of minimum mass \cite{BK}.

Apparently,  a set of   ``monopole"  states
   become massless
as  the bare  quark masses are tuned to a common value,    $ m_i \to m$, at which point
 the low-energy gauge group gets enhanced
from $U(1)\times U(1)$ to $SU(2)\times U(1)$.
For instance,  the  BPS  monopole carrying  magnetic charge
 (1,-1) with respect to two $U(1)$ factors above
has mass proportional to
$ { m_1 - m_2 }$, and appears to become  massless in the limit of
 $SU(2)$  restoration.   Classically   this ``state"  becomes
infinitely extended  in space in such a limit,  and  at the
same time the fields $\phi,  A_i$ degenerate into  trivial
vacuum configuration $\phi(x)=
A_i(x)=0$
  \footnote{This is analogous to   the fate of the 't Hooft -
  Polyakov monopole of the spontaneously broken  $SU(2) \,
  {\stackrel {v} {\longrightarrow}}  \,  U(1) $  theory,   in
 the limit $v\to 0$. }.
More importantly, as the  topological
structure of the theory changes in the $SU(2)$   restoration
limit (from Eq.(\ref{pi1uu})  to Eq.(\ref{pi1suu}), see below)
such a ``massless monopole"  is no longer topologically stable.

\section{Non-abelian Vortices  \label{sec:navort}}
\setcounter{equation}{0}

We will now construct   (BPS)   vortex solutions  in the theory described above and show that
they possess exact   zero modes.

\subsection{Non-abelian Bogomolny  Equations \label{sec:bogom} }

As we have already anticipated,
by restricting ourselves to a particular base submanifold of the Higgs branch of the theory
with four flavors, we are able to deal  with
BPS strings  throughout.     By  gauge and  flavor rotations  the squark VEVS   can be taken   to be of the
form (\ref{qvev}).  Then classically only the two flavors which develop
VEV's will play a role in the vortex solution. Other flavors remain  zero
on the solution, and one can  consider
the squark fields $q^{kA}$ to be $2\times 2$ matrices. Note however that the
additional two flavors are crucial    in the   quantum theory,  in   keeping  the
$SU(2)$ interactions    weakly coupled \footnote{In fact the additional flavors
are important even classically. In the presence of additional flavors
strings can turn into semilocal strings, see \cite{AV} for a review
on semilocal strings. We
shall  not study this issue here.}.

Let us make an ansatz,
\beq 
\label{ansatz}
q^{kA}(x) =\bar{\tilde{q}}^{kA}(x),   \label{squark} \eeq
and a convenient redefinition of the squark fields $q^{kA} \to { 1\o \sqrt2 } q^{kA}. $
The low-energy action   (\ref{qed}) then  reduces ($g_2$ and $g_1$  stand for the $SU(2)$ and $ U(1)$ coupling constants at the scale $\xi$,
respectively)  to
$$
S=\int d^4x \left[\frac1{4g^2_2} \left(F^{a}_{\mu\nu}\right)^2 +
\frac1{4g^2_1}\left(F^{8}_{\mu\nu}\right)^2
+ \left|\nabla_{\mu}
q^{A}\right|^2\right. 
$$
\beq
\left.
+ \frac{g^2_2}{8}\left(\bar{q}_A\tau^a q^A\right)^2+
\frac{g^2_1}{24}\left(\bar{q}_A q^A - 2\xi
 \right)^2
\right],
\label{le}
\eeq
where we have set the   adjoint scalar fields   to their VEVs  (\ref{avev}).
 The string tension can be  written \`a la
Bogomolny \cite{Bo}
\bea
T &=&\int{d}^2 x   \left (    \sum_{a=1}^3  \left[\frac1{2g_2}F^{(a)}_{ij } \pm
     \frac{g_2}{4}
\Big(\bar{q}_A\tau^a q^A\right)
\epsilon_{ij} \right]^2+
\left[\frac1{2g_1}F^{(8)}_{ij} \pm
     \frac{g_1 }{4\sqrt{3}}
\left(|q^A|^2-2\xi \right)
\epsilon_{ij }\right]^2   \non  \\
&+&  \frac{1}{2} \left|\nabla_i \,q^A \pm   i   \epsilon_{ij}
\nabla_j\, q^A\right|^2
\pm
\frac{\xi}{  \sqrt{3}}\tilde{F}^{(8)}
\Big)
\label{bog}
\eea
 where $ \tilde{F}^{(8)} \equiv     { 1\o 2}  \epsilon_{ij}  {F}^{(8)}_{ij }$,
leading to the following first order equations for strings
\bea
&&  \frac1{2g_2 }F^{(a)}_{ij}+
     \frac{g_2}{4}\varepsilon
\left(\bar{q}_A\tau^a q^A\right)   \epsilon_{ij}=0, \qquad  a=1,2,3;
\non \\
&&   \frac1{2g_1}F^{(8)}_{ij}+  
     \frac{g_1}{4\sqrt{3}}\varepsilon
\left(|q^A|^2-2\xi \right)\epsilon_{ij}=0;
\non \\
 &&    \nabla_i \,q^A +i \varepsilon\epsilon_{ij}
\nabla_j\, q^A=0, \qquad    A=1,2,\ldots, N_f .
\label{F38}  \eea
Here $\varepsilon = \pm$ is the sign of the total flux specified below.

The $U(1) \times U(1)$   string solutions  found    in  the case of unequal quark masses   \cite{MY}    can be readily   recognized as
particular solutions of these equations.  To construct them we further restrict  the
gauge field $A_{\mu}^a$ to the single  color component $A_{\mu}^3$   (by setting $A_{\mu}^1 = A_{\mu}^2=0 $), and consider only squark
fields  of the   $2 \times 2$     color-flavor diagonal form:
\beq 
q^{kA}(x)=\bar{\tilde{q}}^{kA}(x)   \ne 0,  \qquad  {\rm for} \quad   k=A   =1,2,
\eeq
by setting all other components to zero.
For unequal masses the relevant topological classification was
\beq
\label{pi1uu}
\pi_1( \frac{U(1)\times U(1)}{{\mathbb Z}_2} )= {\mathbb Z}^{2}
\eeq
and   the  allowed  strings formed a lattice labeled by two integer winding numbers.
In particular, assume that the first flavor winds $n$ times while
the second flavor winds $k$ times and look for solutions of
(\ref{F38}) using the following ansatz \footnote{We use a notation slightly different  from the one used in \cite{MY}:
 $\phi_1(r)$ instead of $\varphi_u(r)$;
  $\phi_2(r)$ instead of $\varphi_d(r)$.   The cylindrical coordinates are here denoted as  $(z, r, \varphi$), the vortex center extending  along the $z$
axis.  }
$$
q^{kA}(x) =\left(
\begin{array}{cc}
  e^{ i \, n\,\varphi  }\phi_1(r) & 0  \\
  0 &  e^{i \, k \,  \varphi }\phi_2(r) \\
  \end{array}\right),
$$
$$
A^3_{i}(x) = -\varepsilon\epsilon_{ij}\,\frac{x_j}{r^2}\
\left((n-k)-f_3(r)\right),\;
$$
\beq
\label{sol}
A^{8}_{i}(x) = -\sqrt{3}\ \varepsilon\epsilon_{ij}\,\frac{x_j}{r^2}\
\left((n+k)-f_8(r)\right)\,
\eeq
where $(r, \varphi) $ are polar coordinates in the (1,2) plane while
the profile functions $\phi_1$, $\phi_2$ for scalar fields and
$f_3$, $f_8$ for gauge fields depend only on $r$.

With this ansatz the first-order equations (\ref{F38}) take the form \cite{MY}
$$
r\frac{\rm d}{{\rm d}r}\,\phi_1 (r)- \frac12\left(f_8(r)
+  f_3(r)\right)\phi_1 (r) =\ 0,
$$
$$
r\frac{\rm d}{{\rm d}r}\,\phi_2 (r)- \frac12\left(f_8(r)
-  f_3(r)\right)\phi_2 (r) =\ 0,
$$
$$
-\frac1r\,\frac{\rm d}{{\rm d}r} f_8(r)+\frac{g^2_1}{6}\,
\left(\phi_1(r)^2 +\phi_2(r)^2-2\xi\right)\ =\ 0,
$$
\beq
-\frac1r\,\frac{\rm d}{{\rm d}r} f_3(r)+\frac{g^2_2}{2}\,
\left(\phi_1(r)^2 -\phi_2(r)^2\right)\ =\ 0.
\label{foe}
\eeq
The profile functions in these equations
are determined by  the following boundary conditions
$$
f_3(0) = \varepsilon_{n,k}\left( n-k\right),
\quad f_8(0)=\varepsilon_{n,k}\,\left( n+k\right), \;
$$
\beq
f_3(\infty)=0, \quad   f_8(\infty)=0
\label{fbc}
\eeq
for the gauge fields, and the requirement that the squark fields be everywhere regular.
The behavior of the latter  at $r=\infty$,
\beq
\label{phibc}
\phi_1 (\infty)=\sqrt{\xi},\quad   \phi_2 (\infty)=\sqrt{\xi}
\eeq
and  that   at $r=0$  (e.g.,  $\phi_1(0)=0,$   if  $n \ne 0, \,\, k=0$),  follow from these requirements.
Here the sign of the string flux is
\beq
\label{sign}
\varepsilon = \varepsilon_{n,k} = \frac{n+ k}{|n+k|} =
{\rm sign}(n+ k) = \pm 1.
\eeq

The tension of a $(n,k)$-string
for the case of equal quark masses is determined  by the
flux of the $A^8_{\mu}$ gauge field
alone    and is given by
\beq
\label{ten}
T_{n,k}=\,2\pi \, \xi \, |n+ k| .
\eeq
Note that $(1,0)$  and $(0,1)$-strings are exactly degenerate.

Note  also  that  ${\tilde F}^{(3)}$ does not
enter the central charge of the \ntwo algebra and so does not affect the string tension.
The stability of the string in this case is due to the $U(1)$
factor of the $SU(2)\times U(1)$ low-energy group only.

The  equations  (\ref{F38}) represent  a nonabelian  generalization of the Bogomolny
equations for the ANO string \cite{Bo}.
For a generic $(n,k)$-string equations (\ref{foe}) do not
 reduce to the standard Bogomolny
equations. For instance,   for  the $(1,1)$-string these equations reduce to
two Bogomolny equations while for the $(1,0)$ and $(0,1)$ strings
they do not.

\begin{figure}[h]
\begin{center}
\epsfig{file=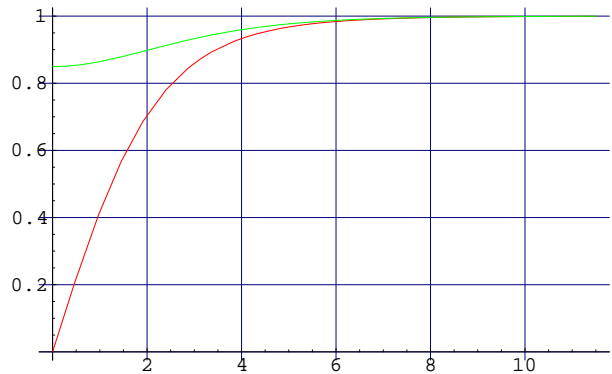,  width=8cm}
\end{center}
\caption{ Vortex profile functions $\phi_1(r)$ and $\phi_2(r)$ of
the $(1,0)$-string. Note $\phi_1(0)=0$.   }
\label{Vort1}
\end{figure}

\begin{figure}[h]
\begin{center}
\epsfig{file=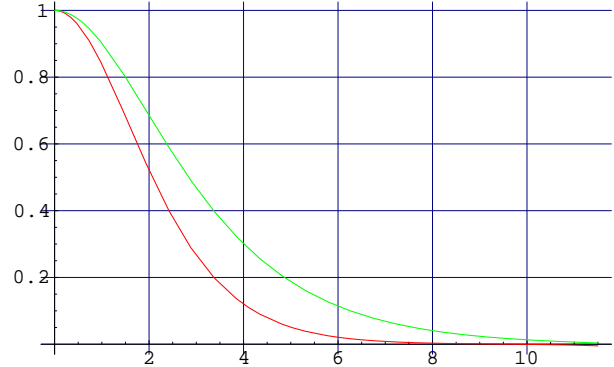,  width=8cm}
\end{center}
\caption{ The profile functions $f_3(r)$ (lower curve) and
$f_8(r)$ (upper curve) for the  $(1,0)$-string.    }
\label{Vort2} \end{figure}

The numerical solution for the
``elementary'' $(1,0)$ string  is  shown in Fig. \ref{Vort1}, Fig. \ref{Vort2}.
The $(0,1)$ string is  obtained  by the replacement,  $\phi_1 \leftrightarrow \phi_2;$ 
$f_3 \leftrightarrow  - f_3.$

The charges of $(n,k)$-strings can be plotted on the Cartan plane of
the $SU(3)$ algebra.
 We shall   use the convention of labeling
the flux of a given  string by the magnetic charge of the monopole which
produces this flux and must be attached to each end.
 This is possible since both string fluxes and monopole
charges are elements of the group $\pi_1(U(1)^2) =
{\bf Z}^{2}$. This convention is convenient because specifying
the flux of a given string automatically fixes the charge of
the monopole that it confines.

\begin{figure}[h]
\begin{center}
\epsfig{file=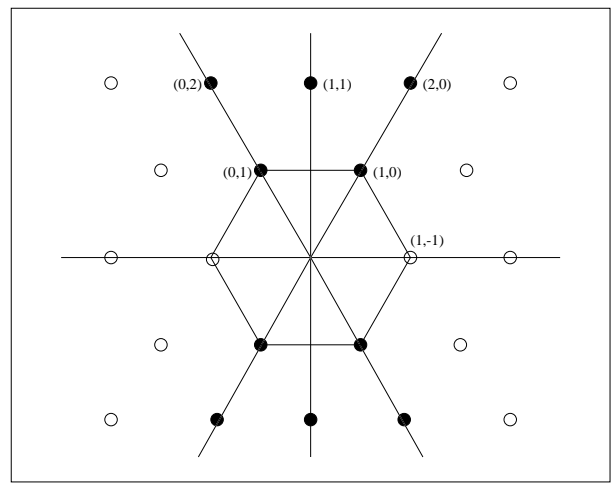,  width=8cm}
\end{center}
\caption{ Lattice of $(n,k)$   vortices.    }
\label{fi:lattice}
\end{figure}

 Our strings are formed by the condensation of squarks
 which have electric charges equal to the  weights of $SU(3)$ algebra.
The Dirac quantization condition tells us that
the lattice of  $(n,k)$-strings is
 formed  by roots of the   $SU(3)$ algebra \cite{MY}.
The lattice of  $(n,k)$-strings is shown in
Fig.~\ref{fi:lattice}. Two strings  $(1,0)$ and $(0,1)$ are the
``elementary'' or ``minimal"   BPS strings. If we plot two lines along charges of these
``elementary'' strings (see Fig.~\ref{fi:lattice}) they divide the
lattice into four sectors. It turns out \cite{MY} that the strings
in the upper and lower sectors, which are labeled by black circles in
Fig.~\ref{fi:lattice},
are BPS but they are marginally
unstable at real quark mass ratios. Instead, strings in
the right and left sectors, which are labeled in
Fig.~\ref{fi:lattice} by white circles,
are bound states of the ``elementary''
ones but they are not BPS.

\subsection{Minimal vortex of generic   orientation: $S^2$  zero modes \label{sec:zeromodes}}

Actually,  the relevant
  homotopy group  here  is  
\beq
\label{pi1suu}
\pi_1( \frac{SU(2)\times U(1)}{{\mathbb Z}_2} )= {\mathbb Z},
\eeq
instead of (\ref{pi1uu}), as we are working with the case of equal quark masses where  
the low-energy gauge group is   $\frac{SU(2)\times U(1)}{{\mathbb Z}_2}$.
The generator of the fundamental group is a loop which encircles the $U(1)/ Z_2$ once \cite{five}, and thus to calculate
 the tension of a string, or to determine whether it
is stable, it suffices to simply count the winding number around this circle.
 This means that the lattice of $(n,k)$-strings reduces to a tower
labeled by one integer $(n+k)$.
For instance, the $(1,-1)$-string becomes
completely unstable    as it winds forward once and
then backward once, and so there is no net winding and so
no topological charge. On the restored $SU(2)$ group manifold
it is also trivial, as it goes half way around the equator and then goes back.  The $(2,0)$ string goes 
all of the way around the $SU(2)$ equator, making a contractible
loop, but is stable because it wraps the $U(1)/Z _2$ twice (it wraps the original $U(1)$ once).

\begin{figure}[h]
\begin{center}
\epsfig{file=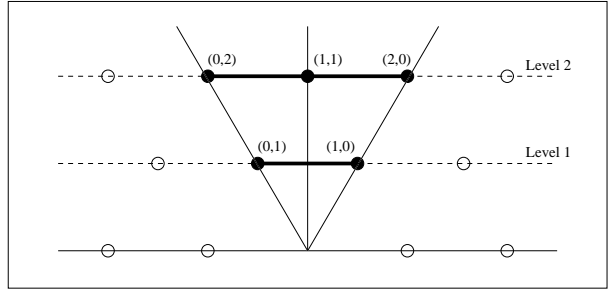,  width=8cm}
\end{center}
\caption{ Reduced lattice of ${\mathbb Z}$   vortices.    }
\label{newlattice}
\end{figure}

On the other hand, the
$(1,0)$ and $(0,1)$ strings cannot be shrunk because they correspond
to a half circle along the equator. They have
the same tension (see (\ref{ten})) for equal quark masses and
thus apparently  belong to doublet of an   $SU(2)$.

In general non-BPS strings on the $(n,k)$-lattice (see  Fig. 3)
become  unstable as they have tensions above their BPS bounds and
we are left with $|n+k|+1$ BPS strings at each winding number
$n+k$.
The reduction of the string lattice is
illustrated  Fig.~\ref{newlattice}.

\begin{figure}[h]
\begin{center}
\epsfig{file=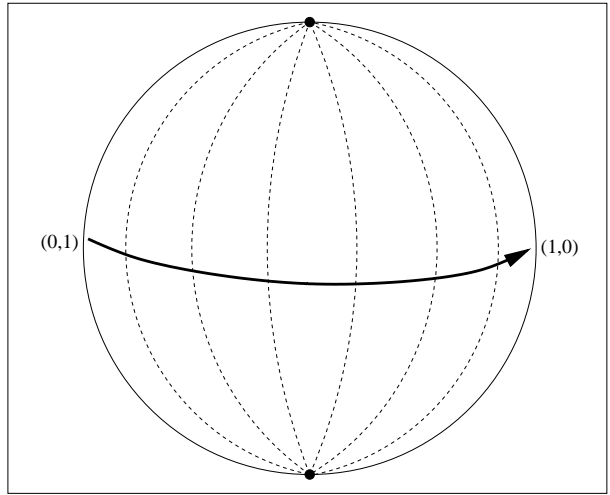,  width=8cm}
\end{center}
\caption{Interpolating between the $(1,0)$-string  and  $(0,1)$-string. }
\label{contin}
\end{figure}

Most importantly,   this suggests that there be  a continuously infinite  number of vortices of minimum winding
and  with the same tension,
\beq
\label{miniten}
T_{1}=\,2 \,  \pi  \,\xi
\eeq
of which  the $(1,0)$
and  $(0,1)$  vortices discussed above are just two
particular cases    (Fig. \ref{contin}).
Below we show that this  is indeed   correct,   by
a continuous deformation of the $(1,0)$-string solution transforming
it into a $(0,1)$-string. This deformation leaves the string tension
unchanged and  therefore  corresponds to an orientational  zero mode.

First let us separate physical variables from the gauge phases eaten by
the Higgs mechanism in the quark fields.
 To do so we use the following parametrization
of the $2\times 2$ quark matrix
\beq
\label{param}
q^{kA}=U_{U(1)}U_{SU(2)}\left(q^0 + \tau^a q^a\right).
\eeq
Here $U_{U(1)}$ and $U_{SU(2)}$ are matrices from the $U(1)$ and $SU(2)$
gauge factors respectively while $q^0(x)$ and $q^a(x)$ are real.
 The parametrization
(\ref{param}) represents eight real variables $q^{kA}$ in terms of
3+1=4 gauge phases eaten by the Higgs mechanism and four physical
variables $q^0$ and $q^a$. 
In particular,
(\ref{qvev}) corresponds to
\beq
\label{vevs}
<q^0>=\sqrt{\xi},\qquad   <q^a>=0.
\eeq

Now let us fix the unitary  gauge (at least globally, which is enough for our
purposes) by imposing the condition that
squark VEV's are given precisely by (\ref{vevs}) and so all gauge phases are zero.
Now transform  the $(1,0)$-string solution (\ref{sol}) into
unitary gauge, which corresponds to the singular gauge in which the
string flux comes from the singularity of the gauge potential at zero.
In this gauge the solution (\ref{sol}) for the $(1,0)$-string
takes the form
$$
q^{kA}=\left(
\begin{array}{cc}
  \phi_1(r) & 0  \\
  0 &  \phi_2(r) \\
  \end{array}\right),
$$
\beq
A^3_{i}(x) = \epsilon_{ij}\,\frac{x_j}{r^2}\,
f_3(r),
\qquad
A^{8}_{i}(x) = \sqrt{3}\ \epsilon_{ij}\,\frac{x_j}{r^2}\,
f_8(r).
\label{s10}  \eeq
Note   that  a global diagonal subgroup in the product
of gauge and flavor symmetries $SU(2)_{C}\times SU(2)_{F}$     is
not broken by the squark VEV. Namely,
\beq
\label{cf}
U<q>U^{-1}=<q>,
\eeq
where $U$ is a global rotation in $SU(2)$ while the squark VEV matrix
is given by (\ref{qvev}). We call this unbroken group   $SU(2)_{C+F}$.

Now let us apply this global rotation to the $(1,0)$ string solution
(\ref{s10}). We find
$$
q^{kA}=U\left(
\begin{array}{cc}
  \phi_1(r) & 0  \\
  0 &  \phi_2(r) \\
  \end{array}\right)U^{-1},
$$
\beq
A_{i}(x) = \frac12\,U\tau^3 U^{-1}\epsilon_{ij}\,\frac{x_j}{r^2}\,
f_3(r),
\qquad   \label{sna}
A^{8}_{i}(x) = \sqrt{3}\ \epsilon_{ij}\,\frac{x_j}{r^2}\,
f_8(r),
\eeq
where we use a matrix notation for the $SU(2)$ gauge field, $A_{\mu}=
A_{\mu}^a\tau^a/2$.
Using the representation \footnote{Explicitly,  if  $n^a=(\sin \alpha \cos \beta,  \sin \alpha  \sin \beta , \cos \alpha)$, the rotation matrix is given
by   $U = \exp{ -i \beta\, {\tau_3 /  2} } \, \exp{ -i \alpha  \, {\tau_2 / 2} } $. }
\beq
\label{na}
U\tau^3 U^{-1}=n^a\tau^a,
\eeq
where $n^a$ is a unit vector on $S^2$, $n^2=1$,   we can rewrite the
$SU(2)$ gauge potential of (\ref{sna}) in the form
\beq
\label{Aa}
A_{i}(x) = \frac12\,n^a\tau^a\epsilon_{ij}\,\frac{x_j}{r^2}\,
f_3(r),
\eeq
revealing that now the $SU(2)$ flux of the string is directed along an arbitrary
vector $n^a$.  It is easy to see that the rotated string
(\ref{sna}) is a solution of nonabelian first order equations
(\ref{F38}).  

 Since  the $SU(2)_{C+F}$  symmetry   is not broken by squark VEV's it is physical and does not
correspond to any of the gauge rotations eaten by the Higgs mechanism. To see
this explicitly let us rewrite the quark field of our solution
(\ref{sna}) using the parametrization (\ref{param}). We get
$$
U_{U(1)}=I,\; U_{SU(2)}=I,
$$
\beq
\label{q0qa}
q^0(x)=\frac12(\phi_1+\phi_2),\qquad  q^a(x)  =n^a\frac12(\phi_1-\phi_2).
\eeq
We see that all gauge phases are zero while physical variables acquire
an $n$-dependence.
Clearly the   solution (\ref{sna}) interpolates between $(1,0)$ and $(0,1)$
strings. In particular it gives a $(1,0)$-string for $n=(0,0,1)$ and
a $(0,1)$-string for $n=(0,0,-1)$.

  The  $SU(2)_{C+F}$  symmetry is exact and the tension of the string solution (\ref{sna}) is independent of   $n^a$:
\beq
\label{eten}
T=2 \,\pi \, \xi,
\eeq
see (\ref{ten}).    However,  an explicit vortex solution breaks the exact $SU(2)_{C+F}$ as
\beq  SU(2)_{C+F} \to U(1):
\eeq
 the two angles associated with vector $n^a$ - two orientational bosonic zero modes of the string - parametrize    the
quotient space $SU(2)/U(1) \sim {\bf CP}^1 \sim S^2$.

In the regular gauge,
the  minimal  nonabelian vortex of generic orientation  (\ref{sna})  takes  the form
\bea
&&    q^{kA}=U\left(
\begin{array}{cc}
 e^{  i \,  \varphi}   \phi_1(r) & 0  \\
  0 &  \phi_2(r) \\
  \end{array}\right)U^{-1}=   e^{ \frac{i}{2} \, \varphi \,  (1+n^a\tau^a)} \,  U\left(
\begin{array}{cc}
  \phi_1(r) & 0  \\
  0 &  \phi_2(r) \\
  \end{array}\right)U^{-1},
\non \\
&&   {\bf A}_{i}(x) = U  [- {\tau^3\o  2} \, \epsilon_{ij}\,\frac{x_j}{r^2}\,
[1-f_3(r)] ]   U^{-1} = -\frac12\,n^a \tau^a\epsilon_{ij}\,\frac{x_j}{r^2}\,
[1-f_3(r)],
\non \\
&&{\hbox {\rm and }} \non \\
\label{rna}
&&   A^{8}_{i}(x) = -\sqrt{3}\ \epsilon_{ij}\,\frac{x_j}{r^2}\,
[1-f_8(r)],
\eea
where $U$ is given by  Eq.~(\ref{na}) and the profile functions are solutions of Eq.(\ref{foe})  for $(n,k)=(1,0)$.
In this gauge it is   particularly clear that  this solution smoothly interpolates between  the $(1,0)$ and
$(0,1)$ solutions:   if $n=(0,0,1)$ the
first flavor squark  winds at infinity while for $n=(0,0,-1)$ the second flavor    does.

To further convince ourselves    that the rotation considered above corresponds
to physical zero modes we can construct a gauge invariant operator
which has $n^a$-dependence on our solution. One example is
\beq
\label{O}
O(x)^A_B= \bar{q}_B \, q^A (x),
\eeq
which is a matrix in flavor indices. Inserting the solution (\ref{rna}) this
operator reads
\beq
\label{Osol}
O(r)=\frac12(\phi_1^2+\phi_2^2)(r)+n^a\tau^a\,\frac12(\phi_1^2-\phi_2^2)(r).
\eeq
We see that  $O(x)$ is a gauge invariant operator which has $n^a$-dependence
localized near the string axis where $(\phi_1^2-\phi_2^2)$ is non-zero.

As we have already mentioned the central charge of the \ntwo algebra
reduces to the ${\tilde F^{8}} $ component of the flux
\beq
\label{8flux}
\frac1{\sqrt{3}}\int d^2 x  \, {\tilde F^{8}} = 2 \, \pi.
\eeq
 If we define a gauge
 invariant flux $\int d^2 x \, F^{a*} \, \Phi^a$ it  reduces to the one
in (\ref{8flux})
so the $SU(2)$ component of the flux does not enter\footnote{We can
define an $SU(2)$ flux by constructing the operator
$2\int d^2 x <\bar{q}>F^{*}<q>/\xi$ which is invariant under global gauge
transformations. This flux is a matrix in flavor indices and on the
string solution (\ref{rna}) reads $2\pi n^a(\tau^a)^A_B$.}. Still as we see
from (\ref{Osol}) there are gauge invariant quantities which acquire
$n$-dependence.

\subsection {Non-abelian monopoles as a multiplet of the  unbroken   dual group}

  In a sense,  the result of the preceeding subsection  solves, albeit indirectly,  the longstanding   ``existence problem"  for the nonabelian
monopoles discussed in the   literature \cite{CDyons}.   In our model (with $m \gg \Lambda$), the monopoles  generated by the symmetry breaking
\beq    SU(3) \Longrightarrow   \frac{SU(2)\times U(1)}{{\mathbb Z}_2}
\eeq
are massive  solitonlike states, which can appear as the   sources of  our  vortices.   The existence of the minimum
vortices with  generic orientation zero modes,   which allows us to  interpolate between the $(1,0)$ and   $(0,1)$-string solutions via $SU(2)$   rotations,    implies the
existence  of the monopoles which
behave truely as  a doublet ($(1,0)$ and  $(0,1)$)  of an  $SU(2)$  group.

How has the ``no-go" theorem of  \cite {CDyons}   been avoided?  First of all,   these monopoles are non-local, finite-energy  soliton states.  The
transformations among  these configurations  must be  in the {\it dual} $SU(2)$  group, and not under the original,
``electric"
$SU(2)$ subgroup.   Topological obstructions  found in attempting to define  globally the ``electric" $SU(2)$  group  in the monopole sector,  do not
apply to the dual group rotations, which are seen here indirectly as a consequence on the sources due to  the global $SU(2)_{C+F}$  actions  on the  {\it
vortices}.

Secondly, the existence of a massless flavors  is fundamental to all of this.  In fact,  the orientation zero modes of the vortices  are generated by the
color-flavor  diagonal $SU(2)_{C+F} $   which is an exact symmetry of the system.    The fact that the dual of  a gauge group involves
the flavor group in some way, may  appear  surprising, but  is not. In fact, it is one of the characteristic  features of Seiberg's duality in ${\cal N}=1$
models. In MQCD it is yet less surprising as pairs of color and flavor branes fuse together in our vacua.  

The fundamental importance of the massless flavors  in generating truely nonabelian monopoles has already been   noted in \cite{BK},
where  it was pointed out that    because of renormalization effects   only  in a theory  with a sufficient number of massless quark flavors does  an
unbroken (dual)  gauge  group remain  exact at low energies.  Otherwise,   the semiclassical pattern of  symmetry breaking   has little to do with the
true symmetry of the system.  If the ``unbroken group" is to be dynamically broken further at low energies,   the degenerate
multiplets of monopoles  found  in the semiclassical approximation mean  simply the presence of an  {\it approximately }   degenerate
set of monopoles.  This is what occurs   in a generic point of space of vacua in
\ntwo  gauge theories.

Coming back  to our  model,   the original {\it local }   $SU(2)\times U(1)$  groups are completely broken by the squark VEVS  at the scale
$\xi= \sqrt{\mu \, m} \ll m $:   the theory is in a Higgs phase.  The dual   $SU(2)$   theory must be in confinement phase. The (massive) doublet
monopoles  are confined.   These conclusions are perfectly consistent with the result of the section \ref{sec:world}    where we study the
dynamics of the fluctuation  of the    $S^2$  zero modes.

\section {Nonabelian Vortices  in  $SU(N)$ Gauge  Theory   }

It is not difficult to generalize   the whole discussion  to the
more general case in which the unbroken gauge group is
\beq        SU(K ) \times  U(1).
\eeq
 For instance,  in the semiclassical vacuum  of the $SU(N)$  theory where  the adjoint scalar  has a   VEV of the form
\beq  \bra \Phi  \ket  = {1\over\sqrt{2}} \, {\rm diag} \, (-m, -m, \ldots, -m,  (N-1)\,  m),  \qquad  m \gg  \Lambda,
\label{N-1vac}\eeq
the gauge  symmetry is broken as
\beq        SU(N)  \to   SU(N-1)   \times  U(1).
\label{N-1th}\eeq
The unbroken gauge group remains weakly coupled at all scales if we
take  the number of flavors to be
\beq    2 \, N  >    N_f \ge   2\, (N-1)
\label{N-1fl}\eeq
so  that the semiclassical analysis is valid at all scales.

The construction of Sections \ref{sec:bogom} and \ref{sec:zeromodes}    can be straightforwardly generalized to these more general cases
with  unbroken  $SU(K ) \times  U(1) $   group.   For concreteness, the following equations will refer to the system Eq.~(\ref{N-1th}), hence
$K =N-1$.  
With the ansatz (\ref{squark}) and after rescaling the squark fields as \beq q^{kA} \to { 1\o \sqrt {2}}  \,  q^{kA},\eeq
the action (\ref{le})  takes the form
\beq
S=\int d^4x \left[ \frac1{4g^2}(F_{\mu\nu}^{(a)})^2
  +\frac1{4e^2}(F_{\mu\nu}^{(0)})^2
  +\left|\nabla_{\mu}q\right|^2
  +\frac{g^2}{2}(\bar{q} \, t^a q)^2
  +\frac{e^2}{4 K (K+1)  }(\bar{q}q-  K  \,  \xi)^2  \right],
\label{SUract}\eeq
where  the index   $a$  runs   over $1,2,\ldots,  K^2-1,$
   $ \xi =  \const \,  \sqrt {\mu  \, m }$   and $g$ and $e$  are the $SU(N-1)$ and  $U(1)$  coupling constants, respectively.     The covariant derivative is
defined by
\beq  \nabla_{\mu}=\partial_{\mu}-iA_{\mu}^at^a-i A_{\mu} \, t^0,   \qquad    t^0=  { 1\o  \sqrt { 2 K ( 1 +K)}  } \pmatrix {  {\mathbf 1}_{K \times
K}   & 0 \cr  0 &  -K},
\eeq
$t^a$
being  the generators of
$SU(K)$
in the fundamental representation.

In  the sequel, we shall rewrite the abelian part in terms of
\beq {\tilde e} \equiv  { e \o  \sqrt { 2 K ( 1 +K)}  };    \qquad  {\tilde A_i  } \equiv   { e   \o  {\tilde e} }   A_i \eeq
(and subsequently drop   the tildes)
to simplify the equations somewhat.   The net effect is  a formal replacement
$  \frac{e^2}{4 K (K+1)  } \to \frac{e^2}{2 },
$   in Eq.(\ref{SUract}).

The vortex tension has the form
\bea
T &=& \!\!\int{d}^2 x   \left (    \sum_{a=1}^{K^2-1}  \left[\frac1{2g }F^{(a)}_{ij } \pm
     \frac{g}{2}
\Big( \bar{q}_A \, t^a q^A\right)
\epsilon_{ij} \right]^2   \non  \\   &+&\!\!
\left[\frac1{2  e}F^{(0)}_{ij} \pm
    { e \o 2 }
\left(|q^A|^2-  K  \, \xi \right)
\epsilon_{ij }\right]^2   \non  \\
&+& \!\!  \frac{1}{2} \left|\nabla_i \,q^A  \pm  i \epsilon_{ij}
\nabla_j\, q^A\right|^2
\pm
K     \,\xi  \,  \tilde{F}^{(0)}
\Big),
\label{bog2}
\eea
where  $ \tilde{F}^{(0)} \equiv     { 1\o 2}  \epsilon_{ij}  {F}^{(0)}_{ij }.$   A BPS vortex is a solution of the linear   Bogomolny
equations,
\bea
&&  \frac1{2g } F^{(a)}_{ij } +\varepsilon
     \frac{g}{2}
\left(\bar{q}_A \, t^a q^A\right)
\epsilon_{ij} =0,      \qquad  a=1,2,\ldots, K^2-1,
  \non \\
&&   \frac1{2  e}F^{(0)}_{ij} + \varepsilon
    { e \o 2}
\left(|q^A|^2-  K\, \xi \right)
\epsilon_{ij }=0;
\non \\
 &&    \nabla_i \,q^A +i \varepsilon\epsilon_{ij}
\nabla_j\, q^A=0, \qquad    A=1,2,\ldots, N_f,
\label{F38sur}  \eea
where $\varepsilon = \pm$   is the sign of total flux.

\subsection { Unbroken $SU(3)$}

Let us now consider  the specific
case with $N=4$     (unbroken $SU(3) \times U(1)$   group). 
Three particular solutions of these equations can be found by keeping
$A_{\mu}^3, A_{\mu}^8$
and
$A_{\mu}^{(0)},$
and by setting all other components to zero.
The squark fields are   labeled
by three integers:
$n,k,p$.
These
correspond to the squark winding numbers:
\beq
 q^{kA}=    \left(
 \begin{array}{ccc}
   e^{i \, n \,\varphi } \phi_1(r) & 0 &0 \\
   0 & e^{i \, k \, \varphi}\phi_2(r) & 0 \\
   0 & 0 & e^{i \, p \,  \varphi   }\phi_3(r) \\
   \end{array}\right),
\eeq
with the conditions
\beq
\phi_1,\phi_2,\phi_3 \to { \sqrt{\xi}  }, \qquad   r\to \infty.
\eeq
  As before,  the only relevant  color (vertical)  and flavor  (horizontal)    components  are shown  above, all other components
are set identically to zero in the vortex solution.
At
$\infty$
we have a pure gauge field
$A_{i} \propto \epsilon_{ij}\,\frac{x_j}{r^2} $.
We find the coefficients by imposing that the
covariant derivatives go to zero.
So we have:
$$
 A^3_{i}(x) = -\epsilon_{ij}\,\frac{x_j}{r^2}
 \Big((n-k)-f_3(r)\Big),
$$
$$
 A^{8}_{i}(x) = -\frac1{\sqrt{3}} \epsilon_{ij}\,\frac{x_j}{r^2}
 \Big((n+k-2p)-f_8(r)\Big),
$$
\beq
 A_{i}(x) = -\frac1{3}\epsilon_{ij}\,\frac{x_j}{r^2}
 \Big((n+k+p)-f_0(r)\Big).  \label{gauge}
\eeq
 The profile functions should  tend to zero at $r=\infty$, and their values at the origin (vortex center) are dictated by the regularity of the
gauge fields  ($f_3(0)=  n-k,$  etc).  The first order equations for the profile functions are:
$$
 r\frac{\rm d}{{\rm d}r}\,\phi_1 (r)- \Big(\frac12f_3(r)+\frac16f_8(r)
   +\frac13f_0(r)\Big)\phi_1 (r) =\ 0,
 $$
 $$
 r\frac{\rm d}{{\rm d}r}\,\phi_2 (r)- \Big(-\frac12f_3(r)+\frac16f_8(r)
   +\frac13f_0(r) \Big)\phi_2 (r) =\ 0,
 $$
 $$
 r\frac{\rm d}{{\rm d}r}\,\phi_3 (r)- \Big(-\frac13f_8(r)
   +\frac13f_0(r) \Big)\phi_3 (r) =\ 0,
 $$
$$
 -\frac1r\,\frac{\rm d}{{\rm d}r} f_3(r)+g^2\,
 \Big(\frac12\phi_1(r)^2 -\frac12\phi_2(r)^2\Big)\ =\ 0.
 \label{foebis}
 $$
 $$
 -\frac1r\,\frac{\rm d}{{\rm d}r} f_8(r)+g^2\,
 \Big(\frac1{2}\phi_1(r)^2 +\frac1{2}\phi_2(r)^2-
  \phi_3(r)^2   \Big)\ =\ 0,
 $$
\beq
 -\frac1r\,\frac{\rm d}{{\rm d}r} f_0(r)+3e^2\,
 \Big(\phi_1(r)^2 +\phi_2(r)^2+\phi_3(r)^2-3 \, \xi\Big)\ =\ 0.
\label{proflesu3}\eeq
The tension of this vortex is given by  the $U(1)$  flux only,
$$ T_{n,k,p}=2 \, \pi \, \xi \,  | \, n+k+p  \, |. $$

The
$(1,0,0),$
$(0,1,0)$
and
$(0,0,1)$-strings all have the same tension.
In the
$A_i^3$
and
$A_i^8$
plane,
they form a equilateral     triangle that corresponds
to the antifundamental of
$SU(3)$.
It is  possible to go from the
$(1,0,0) \to (0,1,0)$
with the Weyl reflection:
$$f_3 \to -f_3,\qquad    \phi_1 \leftrightarrow \phi_2,$$
 other profile functions being  left invariant.
This corresponds to the global color-flavor
$SU(3)_{C+F}$
rotation:
$$U_{C+F}=\left(
    \begin{array}{ccc}
    &1&\\
    -1&&\\
    &&1\\
    \end{array}\right).$$
To do the transformation
$(1,0,0) \to (0,0,1),$
we use the Weyl reflection:
$$f_3 \to \frac12 f_3-\frac12f_8; \qquad   f_8\to-\frac32f_3-\frac12f_8,  \qquad  \phi_1\leftrightarrow\phi_3. $$
In the
$A^3,A^8$
plane, this transformation is exatly the reflection:
$$ R_{Weyl}=
   \left(\begin{array}{cc}
   \cos{-\frac{\pi}3}&\sin{-\frac{\pi}3}\\
   \sin{-\frac{\pi}3}&-\cos{-\frac{\pi}3}\\
   \end{array}\right).$$
The transformation corresponds to the colour-flavour rotation:
$$ U_{C+F}=\left(
    \begin{array}{ccc}
    &&1\\
    &1&\\
    -1&&\\
    \end{array}\right).$$
It is easy to see that with these tranformations,
equations Eq.~(\ref{proflesu3})     and the asymptotic conditions
are left invariant.

We have found that these three solutions belong to the same
set, and it  is possible to continuously interpole between 
them with a
gauge-flavor rotation. This set is parametrized by the coset
$SU(3)_{C+F}/H$,
where
$H$
is the group left invariant by the vortex solution.
Let us check this for the
$(1,0,0)$
vortex.
It is easily seen  that it is  possible to fix:
$$\phi_2=\phi_3=\phi, \qquad    f_3=f_8=f_{NA}$$
 to  reduce to four the number of  profile functions satisfying:
 $$
 r\frac{\rm d}{{\rm d}r}\,\phi_1 (r)- \Big(\frac23f_{NA}(r)+\frac13f(r)\Big)\phi_1 (r) =\ 0,
 $$
 $$
 r\frac{\rm d}{{\rm d}r}\,\phi (r)- \Big(-\frac13f_{NA}(r)
 +\frac13f(r) \Big)\phi (r) =\ 0,
 $$
 $$
 -\frac1r\,\frac{\rm d}{{\rm d}r} f_{NA}(r)+g^2\,
 \Big(\frac12\phi_1(r)^2 -\frac12\phi(r)^2\Big)\ =\ 0.
 \label{foeN}
 $$
 \beq
\label{profile-functions-(1,0,0)}
 -\frac1r\,\frac{\rm d}{{\rm d}r} f(r)+3e^2\,
 \Big(\phi_1(r)^2 +2\phi(r)^2-3\xi\Big)\ =\ 0.
\eeq
Now it  is possible to see that there is un unbroken subgroup
$SU(2) \times U(1)$.
For the
$(0,0,1)$
vortex
we can put:
$$\phi_1=\phi_2=\phi,\qquad    f_3=0,\,\,\,f_8=-  2f_{NA}.$$
The four equations are the same as Eq.~(\ref{profile-functions-(1,0,0)}),
with
$\phi_1$
replaced by
$\phi_3$.

To summarize,  these vortices  possess  exact  orientation zero modes, due to the fact that
the system has an exact  color-flavor  diagonal
symmetry, $    SU(3)_{C+F}
$.     Since  any particular vortex solution, like
those found above, breaks this symmetry as
\beq   SU(3) \to SU(2) \times  U(1),
\eeq   there actually exist   a continuous family of solutions of the same tension.
The vortices of minimum tension of   a generic orientation in  $SU(3)$   are  constructed   starting  from {\it e.g.,}   the $(1,0,0)$ solution,
by
$ SU(3)_{C+F} $    transformations
\beq
q^{kA}=   U     \,  \left(
 \begin{array}{ccc}
   e^{i   \varphi  }\phi_1(r) & 0 &0 \\
   0 & \phi_2(r) & 0 \\
   0 & 0 &   \phi_2(r) \\  
   \end{array}\right)    U^{\dagger},
\eeq
\beq   A_i  =    U   A^{(1,0,0)}_i   U^{\dagger},
\eeq
where   $A^{(1,0,0)}_i$  stands for the gauge fields  (\ref{gauge})  with  $(n, k, p) =(1,0,0).$     This family of   vortices  are
labeled by the four  real parameters  of
\beq {SU(3) \o  SU(2) \times U(1)}  \sim{\bf CP}^{2}.
\eeq

\subsection{Generalization to K-vacua}

In the case with unbroken $SU(K)$  symmetry, one apparently has
$2K$
profile functions:
\beq  \phi_1,\ldots,\phi_K,\qquad      f_3,  \ldots,  f_{K^2-1}, \quad  f, \eeq
where  $f_{k^2-1}$'s   ($k=2,3,\ldots, K$) correspond to   $K-1$ generators of the  Cartan subalgebra.
The ansatz is:
$$ 
 q^{kA}=\left(
 \begin{array}{ccc}
   e^{i \, n_1\alpha}\phi_1 & 0 &0 \\
   0 & \ddots & 0 \\
   0 & 0 & e^{i \, n_K\alpha}\phi_K \\
   \end{array}\right),
$$
$$
 A^3_{i}(x) = -\epsilon_{ij}\,\frac{x_j}{r^2}
 \Big((n_1-n_2)-f_3\Big),
$$
$$\vdots$$
$$
 A^{K^2 -1}_{i}(x) = -\sqrt{\frac2{K (K-1)}} \epsilon_{ij}\,\frac{x_j}{r^2}
 \Big((n_1+\dots+n_{K-1}-(K-1)n_K)-f_{K^2-1}\Big),
$$
\beq 
 A_{i}(x) = -\frac1{K }\epsilon_{ij}\,\frac{x_j}{r^2}
 \Big((n_1+\dots+n_K)-f\Big).
\eeq

Actually, the solution leaves  an
$SU(K-1)\times U(1)$
symmetry  invariant, as  can be seen from tha fact that   they can be expressed in terms of four
profile functions only, as   in the  $SU(2)$ and  $SU(3)$  cases studied above.  In fact,  
for the
$(0,\dots,0,1)$  
vortex  one can set:
$$  \phi_1=\dots=\phi_{K-1}=\phi,  $$
\beq  f_{3}=\dots=f_{(K-1)^2 -1}=0,\,\,\,f_{K^2 -1}= -  (K-1)f_{NA}   \eeq
reducing the linear equations to the set:
$$ 
 r\frac{\rm d}{{\rm d}r}\,\phi (r)- \Big(-   \frac1  K  f_{NA}(r)+\frac1 K  f(r)\Big)\phi (r) =\ 0,
$$
 $$
 r\frac{\rm d}{{\rm d}r}\,\phi_K  (r)- \Big(  \frac{(K-1)}{K }f_{NA}(r)+\frac1K f(r) \Big)\phi_K (r) =\ 0,
 $$
 $$
 -\frac1r\,\frac{\rm d}{{\rm d}r} f_{NA}(r)+\frac{g^2}2\,
 \Big(\phi_K(r)^2-\phi(r)^2\Big)\ =\ 0,
 $$
 \beq 
 -\frac1r\,\frac{\rm d}{{\rm d}r} f(r)+K  e^2\,
 \Big((K -1) \, \phi(r)^2+\phi_K (r)^2- K  \xi\Big)\ =\ 0.
 \eeq
These equations reduce to Eq.~(\ref{profile-functions-(1,0,0)})   for
$K  =3$.

Considering that the above system arises from  the softly broken \ntwo $SU(N)$  theory with $N_f $ flavors,    $2N-2  \le N_f \le 2 N$,     broken by
the adjoint scalar VEVS as
\beq   SU(N) \to  { SU(N-1) \times U(1) \o  {\mathbb Z}_N},  \eeq
the system  has an exact  $ SU(N-1)_{C+F}$ symmetry, respected both by the adjoint
and squark VEVS    ($K=N-1$  above).    A vortex solution breaks this symmetry to
\beq     SU(N-1)_{C+F}   \to   SU(N-2) \times    U(1)
\eeq
and consequently a continuous $ 2 (N-2)$-parameter family of degenerate vortices exist,  representing the
quotient space,
\beq  {SU(N-1)  \o   SU(N-2) \times    U(1)} \sim  {\bf CP}^{N-2}.
\label {CPN2}\eeq

\section{The effective vortex   world-sheet theory  \label{sec:world}}
\setcounter{equation}{0}

 We  study now   the effective low-energy
theory for orientational collective coordinates  on the string world sheet.  We first restrict ourselves to the  $SU(3)  \to SU(2) \times U(1)$  theory
of Section 2 and Section 3,  coming back to more general cases later.
   We shall study  the bosonic collective coordinates $n^a$ first and then
use the unbroken  supersymmetry to reconstruct  the
fermionic sector.

\subsection{Kinetic term}

 Assume  that the orientational collective coordinates $n^a$
are   slow varying functions of the string world sheet coordinates
$x_n$, $n=0,3$. Then $n^a$ become fields in a (1+1)-dimensional
sigma model on the world sheet. Since   the vector   $n^a$ parametrizes the string zero modes,
 there is no potential term in this sigma model.  Let us work out the kinetic term.

To do so we substitute our solution (it is convenient to use it in the
singular gauge (\ref{sna})) into the action (\ref{le}) assuming now that
the fields acquire a  dependence on coordinates $x_n$ via $n^a(x_n)$.
 However, before
doing this we have to modify our solution. The point is that
our solution was obtained as a $SU(2)_{C+F}$ rotation of the $(1,0)$-string.
Now we make this transformation local (depending on $x_n$). Therefore,
the $n$-components of the gauge potential are no longer zero. We assume the
obvious ansatz for these components
\beq
\label{An}
A_n=-i\,  \de_n U\,U^{-1}\,f(r),
\eeq
where we have introduced a new profile function $f(r)$. It is determined by its
own equation of motion which we will derive below. This function vanishes at infinity
\beq
\label{bcfinfty}
f(\infty)=0,
\eeq
while the boundary condition at $r=0$ will be determined shortly.

The kinetic term for $n^a$ comes from gauge and quark kinetic terms in
(\ref{le}). Using (\ref{sna}) and (\ref{An}) to calculate the
$SU(2)$ gauge
field strength we find
\beq
\label{Fni}
F_{ni}=\frac12\,\de_n n^a \tau^a\epsilon_{ij}\,\frac{x_j}{r^2}\,
f_3[1-f(r)]+ i \,\de_n \, U\,U^{-1}\,\frac{x_i}{r}\frac{d}{dr}f(r).
\eeq
 We see that in order to have a finite contribution coming from
$TrF_{ni}^2$ we have to impose
\beq
\label{bcfzero}
f(0)=1.
\eeq

Now substituting the field strength (\ref{Fni}) into the action
(\ref{le}) and taking into account also kinetic term for quarks we finally
arrive at
\beq
\label{o3}
S^{(1+1)}_{\sigma}=\beta \int d^2 x  \, \frac12 \left(\de \, n^a\right)^2,
\eeq
where the integration goes over world sheet coordinates $x_n$
while the coupling constant $\beta$ is given by
$$
\beta= \frac{2\pi}{g_2^2} \int_0^{\infty}
r \, dr\left\{\left(\frac{d}{dr}f(r)\right)^2
+\frac1{r^2}f_3^2(1-f)^2+\right.
$$
\beq
\left.
+   g_2^2\left[\frac12 f^2(\phi_1^2+\phi_2^2)
+(1-f)(\phi_1-\phi_2)^2\right]\right\}.
\label{beta}  \eeq

\begin{figure}[h]
\begin{center}
\epsfig{file=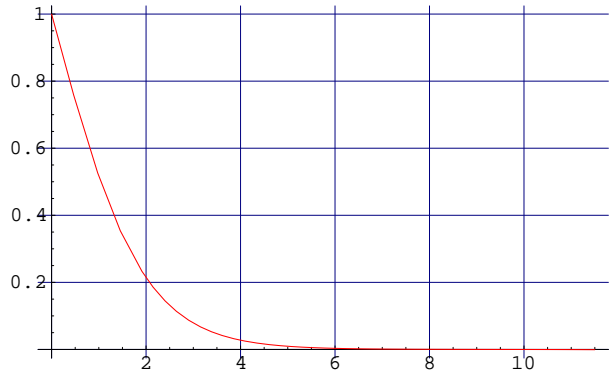,  width=8cm}
\end{center}
\caption{ The  profile function $f(r)$.  }
\label{Vort3}
\end{figure}

We see that the effective world sheet theory for the string orientational
zero mode is given by an $O(3)$ sigma model.  The symmetry group of
this sigma model  is nothing but  global $SU(2)_{C+F}$     whose 3-dimensional representation acts as the group of orientation preserving isometries on the
target space, {\bf CP}${}^1$.
 The coupling constant of this sigma model is
determined by the minimum of action  (\ref{beta}) for the function
$f$.  A numerical solution for the profile function $f(r)$ is given in
Fig.~\ref{Vort3}.
Note   that the function $f$ satisfies a second order equation because,
once  we allow the  dependence of $n^a$ on world sheet coordinates,     the vortex   is no longer BPS saturated. The emergence
of new profile functions which determine the kinetic terms in the
effective world volume theory of a string or domain wall was
observed earlier in \cite{SY,SY2}.    

Clearly  (\ref{o3}) describes an effective low-energy theory.
It has higher derivative corrections    in powers of
\beq
\label{hd}
\frac{\de_n} {g_2\sqrt{\xi}},
\eeq
where we use Eq.~(\ref{m3}) to determine masses of
gauge/quark multiplets in our $SU(2)\times U(1)$ low-energy theory.
The sigma model (\ref{o3}) gives a good description at even lower scales,
well below
$ g_2\sqrt{\xi}$ where higher derivative corrections are small.
The scale $g_2\sqrt{\xi}$
determines also the inverse thickness of our string, in other words
the effective sigma model  (\ref{o3}) can be applied
at scales below the inverse thickness of the string
which plays a role of the UV cutoff for  (\ref{o3}).
It is quite natural  that in
the confinement phase the  effective theory below
the inverse thickness of a string
 becomes a two-dimensional  sigma model
on its world sheet.

\subsection{\ntwo     ~ $O(3)$ sigma model in (1+1) dimensions}

An \ntwo supersymmetric theory in four dimensions has eight
supercharges.  The  string  solution of section 4 is 1/2 BPS.
Thus we have four supercharges in the two dimensional
sigma model on the string world sheet, which generate \ntwo or
more precisely (2,2) supersymmetry in 2 dimensions.
We have seen that our effective theory
(\ref{o3}) on the string world sheet is the bosonic part of the
\ntwo supersymmetric $O(3)$ sigma model in two dimensions.  The
physics of this theory is well understood using the mirror
description \cite{HoVa}. Below we briefly review known results
and interpret them in terms of strings in four dimensions.

The action   of this model reads
\beq
\label{so3}
S_{\sigma}^{(1+1)}=\frac{\beta}{2}\int d^2 x \,  d^2 \theta  \, d^2
\bar{\theta}  \, \log{(1+\bar{W}W)},
\eeq
where $W$ is a chiral superfield
\beq
\label{W}
W= w+ \sqrt{2} \, \theta_{\alpha} \, \psi^{\alpha} +\theta^2 F.
\eeq
Here $w$ is a complex bosonic field related to the vector $n^a$ by
the  stereographic projection
\beq  n^3=\frac{1-|w|^2}{1+|w|^2},
\qquad   n^1=2\frac{Re\,w}{1+|w|^2},
\qquad
\label{stereo}
n^2=2\frac{Im\,w}{1+|w|^2},
\eeq
 while $\psi^{\alpha}$ is a complex
fermion field, $\alpha =1,2$.
The bosonic part of the action (\ref{so3}) has a standard
form
\beq
\label{bo3}
S_{\sigma}^{(1+1)bos}=2 \beta   \int d^2 x  \,
\frac{\de_n \bar{w}\,  \de_n w}{(1+\bar{w} \, w)^2},
\eeq
which is identical to the one in (\ref{o3}) upon substitution
(\ref{stereo}).

Classically the $O(3)$ sigma model has a spontaneous breaking of
the $SU(2)_{C+F}$ symmetry and two massless Goldstone bosons.
This means that for a given string the vector $n^a$ is
pointed towards   a particular direction.

However, the quantum physics of the \ntwo sigma model is quite
different. The model is asymptotically free and runs into
a strong coupling regime at low energies. The
renormalized coupling constant   as a function of
the energy scale $E$ is given by
\beq
\label{sigmacoup}
4\pi \beta = 2\log{(\frac{E}{\Lambda_{\sigma}})} +\cdots,
\eeq
where $\Lambda_{\sigma}$ is the scale of the sigma model.
This scale is determined by the condition that the sigma
model coupling $\beta$ at the scale of the sigma model UV
cut-off $g_2\sqrt{\xi}$ is given by the four dimensional
low-energy coupling $g_2$ via (\ref{beta}). Thus,
\beq
\label{lambdasig}
\Lambda^2_{\sigma}\sim \xi e^{-\gamma\frac{8\pi^2}{g_2^2}},
\eeq
where $\gamma$ is the value of the integral in  the action
(\ref{beta}).

The model has instantons which induce chiral symmetry breaking.
Namely, there is a non-zero chiral  fermion bilinear condensate
\cite{NSVZ}
\beq \label{chcon}
<\frac{\bar{\psi}(1+\gamma_5)\psi}{(1+|w|^2)^2}> =\pm  \,  \const
\Lambda_{\sigma},
\eeq
where $\gamma_5=\tau_3$.  The fact that there are two values of chiral condensate
indicates that there are two vacua in the sigma model.

The physics of the model becomes more transparent in the mirror
description.  This is a description of the model in terms of the Coulomb gas
of instantons, and is equivalent to a sine-Gordon theory \cite{FFS}.
Explicitly, the model (\ref{so3}) is dual to the \ntwo sine-Gordon
theory  \cite{HoVa}
\beq
\label{sG}
S_{\sigma}^{(1+1)}=\int d^2 x \left[ d^2 \theta  \, d^2
\bar{\theta} \, \frac1{\beta} \, \bar{Y} \, Y +\Lambda_{\sigma} \, d \theta^1 d
\bar{\theta}_2 \cosh{Y}\right].
\eeq
Here the last term is a dual superpotential induced by
instantons, while   $Y$ is a twisted chiral  superfield with the
expansion
\beq
\label{Y}
Y= y+ \sqrt{2} \, \theta^1\bar{\chi}_{1}
+\sqrt{2} \, \bar{\theta}_2\chi^{2} +\cdots.
\eeq

This theory has a mass gap of order of $\Lambda_{\sigma}$,
indicating that there is no spontaneous breaking of $SU(2)_{C+F}$
and no Goldstone bosons. 

\subsection{\ntwo     ~ ${\bf  CP}^{N-2}$  sigma model in (1+1) dimensions}

An analogous conclusion follows in  the more general  case
of an $SU(N)$  theory, Eq.(\ref{N-1vac}), Eq.(\ref{N-1th}), Eq.(\ref{N-1fl}).    The low-energy  action and its vacuum   respect
 a global   $ SU(N-1)_{C+F} $ symmetry,  which is broken however   by an individual  vortex configuration    to $ SU(N-2) \times U(1).$   See 
Eq.(\ref{CPN2}).    We assume
that a consideration analogous to the one given for the  $SU(3)$  theory   leads to an   \ntwo,
\beq  {SU(N-1)  \o   SU(N-2) \times    U(1)} \sim  {\bf CP}^{N-2}
\label {CPN2bis}\eeq
sigma model on the  vortex world sheet.  A study of such systems
\cite{AcVa}  shows that the number of vacua in this sigma
model is $N-1$.

\subsection*{Acknowledgements}

We are grateful to Hitoshi Murayama for  discussions on the homotopy group properties of the
nonabelian monopoles   and   Adam Ritz
for  useful conversations.   A. Y.  would like to  thank
Istituto Nazionale di Fisica Nucleare -- Sezione di Pisa, 
for hospitality.
The work of A.Y. was supported by Russian Foundation
for Basic Research under the grant
No~02-02-17115  and  by INTAS grant No~00-00334.
K. K. thanks Japan Society for the Promotion of Science (Fellow ID S-03034) and N. Sakai  (Tokyo Institute of Technology)
for hospitality.

\end{document}